\documentclass{ws-ijmpd}

\begin{document}

\markboth{Merab Gogberashvili}
{Topological Solution to the Cylindrical Einstein-Maxwell Equations}

%
\catchline{}{}{}{}{}
%

\title{Topological Solution to the Cylindrical Einstein-Maxwell Equations}

\author{Merab GOGBERASHVILI}

\address{Andronikashvili Institute of Physics,
6 Tamarashvili Street, Tbilisi 0177, Georgia \\
Javakhishvili Tbilisi State University,
3 Chavchavadze Avenue, Tbilisi 0128, Georgia \\
California State University,
2345 E. San Ramon Avenue M/S MH37, Fresno 93740, USA\\
(e-mail: gogber@gmail.com)}

\maketitle

\begin{abstract}
New exact solution of the cylindrically symmetric Einstein-Maxwell equations is presented. The solution is singular on the axis of symmetry and at the radial infinity, where sources should be placed. The accepted source at the origin can be interpreted as a charged domain wall shell.
\end{abstract}

\keywords{Einstein-Maxwell equations; Cylindrical waves; Domain
wall shell.}

\vskip1cm

In general relativity cylindrically symmetric wave solutions are well-known\cite{Exact}. The general cylindrical line element that is capable of radiation has the form\cite{General}:
\begin{equation}\label{Ansatz}
ds^2 = e^{2(\gamma - \psi)}\left( dt^2 - dr^2\right) - r^2
e^{-2\psi} d\theta^2 - e^{2\psi}\left( dz+ \omega d\theta\right)^2~,
\end{equation}
where the quantities $\gamma$, $\psi$ and $\omega$ are functions of the time $t$ and the radial coordinate $r$. For the wave solution the functions $\psi$ and $\omega$ represent two dynamical degrees of freedom of gravitational field, the linear and the cross polarizations. The most studied subclass of (\ref{Ansatz}) is the case with $\omega = 0$, which in the vacuum corresponds to the Einstein-Rosen gravitational waves\cite{ER}.

The function $\gamma$ in (\ref{Ansatz}) corresponds to the gravitational energy of the system, the so-called $C$-energy\cite{Thorne},
\begin{equation} \label{C}
C(t,r) = \frac 12 \ln{|g_{tt}g_{zz}|}~,
\end{equation}
which gives the total gravitational energy per unit length between the axis of symmetry and the radius $r$ at time $t$. Being a locally conserved and measurable quantity, $C$-energy is a useful tool for the analysis of cylindrical systems. For example, it can be linked to the occurrence of conical singularities\cite{Xan} and trapped surfaces\cite{Hay} of cylindrically symmetric sources. Note also that the quantity
\begin{equation} \label{Energy}
E_{rad}(t,r) = \frac 12 \left[ \dot C(t,r) - C(t,r)'\right]~,
\end{equation}
where the overdot mean time derivative and the prime stand for derivative with respect of radial coordinate, measures the energy radiated away.

In literature there were considered various effects for the wave line element (\ref{Ansatz}) having analogy with electromagnetic radiation. For example, there was found the Faraday rotation of polarization plane of gravitational waves\cite{Faraday}, including rotation between gravitational and electromagnetic waves\cite{Park}; Connections with the Bessel beams was considered in\cite{Kramer}; It was shown that (\ref{Ansatz}) can represent standing gravitational waves\cite{standing}.

In this paper we obtain new exact wave-like solution of the cylindrical Einstein-Maxwell equations and study properties of its feasible source. We work with the system of Einstein-Maxwell field equations\cite{La-Li}
\begin{eqnarray} \label{E}
R_{\alpha\beta} = 8\pi G T_{\alpha\beta}~, \\
\partial_\nu\left(\sqrt{-g}g^{\nu\alpha}g^{\mu\beta}F_{\alpha\beta}\right) = 0~, \label{M}
\end{eqnarray}
where $G$ is the Newton constant and
\begin{equation}\label{T}
T_{\alpha\beta} = \frac{1}{4\pi}\left(-F_{\alpha\mu}F_{\beta}^{\mu} + \frac 14 g_{\alpha\beta}F_{\mu\nu}F^{\mu\nu}\right)
\end{equation}
is the energy-momentum tensor of electromagnetic field.

The non-zero components of Ricci tensor for (\ref{Ansatz}) are:
\begin{eqnarray} \label{Einstein}
R_{tt} &=& \Box \psi -  2 \dot \psi ^2 - \Box \gamma -\frac{1}{2r^2}e^{4\psi}\dot\omega^2 ~, \nonumber \\
R_{rr} &=& -\Box \psi - 2 \psi'^2 + \Box \gamma  + \frac 2r \gamma'- \frac{1}{2r^2}e^{4\psi}\omega'^2 ~, \nonumber \\
R_{\theta\theta} &=& \omega e^{-2\gamma+4\psi}\left[\Box \omega + \frac 2r \omega' + 4 \left(\dot\omega\dot\psi - \omega'\psi' \right)\right] - \nonumber \\
&-& \left(r^2 - \omega^2 e^{4\psi}\right)e^{-2\gamma}\left[\Box \psi - \frac{1}{2r^2} e^{4\psi} \left(\dot\omega^2 - \omega'^2\right) \right] ~, \nonumber \\
R_{zz} &=& e^{-2\gamma + 4\psi}
\left[\Box \psi - \frac{1}{2r^2}e^{4\psi}\left(\dot\omega^2 -\omega'^2\right)\right] ~, \\
R_{rt} &=& \frac 1r \dot\gamma - 2 \dot\psi\psi' - \frac{1}{2r^2}e^{4\psi}\dot\omega \omega'  ~, \nonumber \\
R_{\theta z} &=& \omega e^{-2\gamma + 4\psi}\left[\Box\psi - \frac{1}{2r^2}e^{4\psi}\left(\dot\omega^2-\omega'^2 \right)\right]+ \nonumber \\
&+& \frac 12 e^{-2\gamma + 4\psi}\left[\Box \omega + \frac 2r \omega' + 4 \left(\dot\omega\dot\psi - \omega'\psi'\right) \right]~, \nonumber
\end{eqnarray}
where
\begin{equation}
\Box= \frac{\partial^2}{\partial t^2} - \frac{\partial^2}{\partial r^2}- \frac 1r \frac{\partial}{\partial r}
\end{equation}
represents the D'Alembert operator in cylindrical coordinates.

In cylindrical symmetric case the gauge can be chosen so that the only non-vanishing components of electromagnetic vector potential are $A_\theta(t,r)$ and $A_z(t,r)$\cite{Thorne,Harrison}. Main results of the paper remain the same if for the simplicity we suppose that only the azimutal component of the vector potential is non-zero, i.e.
\begin{equation} \label{A}
A_t = A_r = A_\theta = 0~, ~~~~~ A_z = A(t,r)~.
\end{equation}
In this case the only non-zero components of the electro-magnetic tensor are:
\begin{equation} \label{F}
F_{tz}=\dot A~, ~~~~~ F_{rz}= A'~,
\end{equation}
and the energy-momentum tensor (\ref{T}) takes the form:
\begin{eqnarray} \label{Tmunu}
T_{tt} &=& T_{rr} = -\frac 12 g^{zz}\left(\dot A^2 + A'^2\right)~, \nonumber \\
T_{\theta\theta} &=& \frac 12 g_{\theta\theta}g^{tt}g^{zz}\left(\dot A^2 - A'^2\right)~, \nonumber \\
T_{zz} &=& \frac 12 g^{tt}\left(\dot A^2 - A'^2\right)~,  \\
T_{tr} &=& - g^{zz}\dot A A'~, \nonumber \\
T_{\theta z} &=& \frac 12 g_{\theta z}g^{tt}g^{zz}\left(\dot A^2 - A'^2\right)~. \nonumber
\end{eqnarray}

Using the expressions of the determinant
\begin{equation} \label{det}
\sqrt{-g} = r e^{2(\gamma - \psi)}~,
\end{equation}
and the inverse metric tensor,
\begin{eqnarray}\label{inverse}
g^{tt} &=& - g^{rr} = e^{-2(\gamma - \psi)}~, \nonumber \\
g^{\theta\theta} &=& - \frac{1}{r^2}e^{2 \psi}~,  \nonumber \\
g^{zz} &=& - e^{- 2\psi} - \frac{\omega^2}{r^2}e^{2\psi}~,  \\
g^{\theta z} &=& \frac{\omega}{r^2} e^{2\psi}~, \nonumber
\end{eqnarray}
it can be shown that the Maxwell equations (\ref{M}) for the {\it ansatze} (\ref{Ansatz}) and (\ref{A}) reduce to:
\begin{eqnarray} \label{Maxwell}
\omega \Box A+2\omega \left( \dot A \dot \psi - A'\psi'\right) + \frac 2r \omega A' + \left( \dot\omega\dot A - \omega'A'\right) = 0 ~, \nonumber \\
\Box A-2 \left( \dot A \dot \psi - A'\psi'\right) + \frac{\omega}{r^2} e^{4\psi} \left( \dot\omega\dot A - \omega'A'\right) = 0~ .
\end{eqnarray}
This system is solved if we put
\begin{eqnarray}
\dot \omega \dot A = \omega' A'~, ~~~~~ \ddot A = A''~, \nonumber \\
\dot \psi = 0~, ~~~~~ \psi' = \frac{1}{2r}~.
\end{eqnarray}
So for our choice of electromagnetic field (\ref{F}) we are left with the static gravitational potential $\psi \sim \ln{r}$. The Einstein equations (\ref{E}) set additional restrictions on unknown functions and the final form of new solution is
\begin{eqnarray} \label{Solution}
A &=& A(t \pm r)~, ~\nonumber \\
\omega &=& const~, \nonumber \\
\psi &=& \frac 12 \ln{r}~, \\
\gamma &=& \frac 14 \ln{r} + \frac 14 C(t\pm r) + \ln{D}~. \nonumber
\end{eqnarray}
Here $D$ is a constant related to the global conicity of the space, the vector potential $A(t\pm r)$ is any function of light-like coordinates $|t\pm r|$ and
\begin{equation}
C(x) = \frac 12\left(1+\omega^2\right)\int_0^x{A(y)'^2 dy}
\end{equation}
is the $C$-energy (\ref{C}) of the system. So the line element (\ref{Ansatz}) for our solution (\ref{Solution}) takes the form:
\begin{equation}\label{Ansatz-2}
ds^2 = \frac{D^2 e^{2 C(t\pm r)}}{\sqrt{r}}\left( dt^2 - dr^2\right) - r d\theta^2 -  r\left( dz+ \omega d\theta\right)^2~.
\end{equation}
This metric corresponds to the space-time of some cylindrically symmetric source having the stationary gravitational potentials $\psi$ and $\omega$, which radiates energy by electromagnetic waves according the law (\ref{Energy}).

In literature there exist several other exact solutions to the cylindrical Einstein-Maxwell equations (\ref{E}) and (\ref{M})\cite{Exact}. Even various method of generating exact solution of this system from a given seed metric were developed\cite{invers}. For example, harmonic solution for the case $\omega = 0$ is presented in\cite{Kramer}; most general static solution, again for $\omega = 0$, was found in\cite{Thorne,Melvin}. To the best my knowledge the line element (\ref{Ansatz-2}), which is the $\omega \ne 0$ generalization of the metric considered in\cite{Rao}, represents the new static-harmonic solution to the Einstein-Maxwell equations.

Physical meaning of (\ref{Ansatz-2}), as of any cylindrically symmetric solutions of the combined Einstein-Maxwell equations (\ref{E}) and (\ref{M}), are fairly well understood. Existing solutions in most cases, as well as (\ref{Ansatz-2}), are singular at $r=0$. Our metric has singularity also at $r \rightarrow \infty$. To avoid singularities it is necessary to introduce sources along the axis of symmetry and at the infinity. To understand what kind of source can be placed at the origin for our solution (\ref{Ansatz-2}) lets consider seed metric by plugging $A(t,r) = C(t,r) = \omega =0$. We notice that the seed metric
\begin{equation} \label{Levi}
ds^2 = \frac{D^2}{\sqrt{r}}\left( dt^2 - dr^2\right) - r d\theta^2 - rdz^2~,
\end{equation}
represents the subclass of the Levi-Civita space-time\cite{Levi} with the value of the parameter (that relates to the energy density of the source) equal to $-1/2$ (the power of $r$ in front of $dt^2$ in (\ref{Levi})). So Lewi-Civitas's mass per unit length $M_{LC}$ for the metric (\ref{Levi}) appears to be negative,
\begin{equation}
M_{LC} = - \frac{1}{2G}~,
\end{equation}
what means that the possible source actually repels particles outside the axis of symmetry rather than attracting them\cite{Thorne,cylinder}.

Problems in identifying sources of the Levi-Civita space-time has been reviewed in\cite{Bonnor}, cylinders\cite{cylinder} and cylindrical shells\cite{shell} were studied as a physical acceptable sources. Since Levi-Chivita metric with negative parameter correspond to a repulsive line mass probably best candidate for the source of our metric (\ref{Ansatz-2}) is the charged domain wall shell\cite{Ar-De}. The space-time (\ref{Ansatz-2}) actually is closed at $r \rightarrow \infty$ and it seems that this metric can give geometry between the two coaxial cylindrical shells, which are located at the origin and at the infinity.

The shells as a source of the Levi–Civita metric made of various types of matter can be studied\cite{Ar-De,Bi-Zo} using thin-wall formalism\cite{Isr}. Since our solution requires multiple shell system (at least two, one at the origin and one at the infinity) we assume that interior region of the shell is not flat and can be described by similar to (\ref{Levi}) metric, but with different parameter $D$. After straightforward calculations using thin-wall formalism\cite{Isr} the energy $\epsilon$ and the components of momentum $p_z, p_\theta$ can be written as:
\begin{eqnarray} \label{e,p}
\epsilon = - \frac{1}{8\pi G r^{3/4}} \left(\frac 1D_+ - \frac 1D_-\right)~, \nonumber \\
p_z = p_\theta = \frac{1}{32\pi G r^{3/4}} \left(\frac 1D_+ - \frac 1D_-\right)~,
\end{eqnarray}
where $D_\pm$ are values of the parameter $D$ inside and outside of the cylindrical shell. From these relations we see that to have positively defined energy of the shell we should require $D_+ > D_-$. Also we notice that the shell is isotropic,
\begin{equation}
p_z = p_\theta = p~,
\end{equation}
and its equation of state is
\begin{equation}
\epsilon = - \frac 14 p~.
\end{equation}
which is similar to the equation of state of a domain walls (for a flat domain wall $\epsilon = -  p$). It is easy to check that the shell obeys the weak energy condition:
\begin{equation}
\epsilon \geq 0~, ~~~~~~ \epsilon + p_z \geq 0~, ~~~~~~ \epsilon + p_\theta \geq 0~,
\end{equation}
the dominant energy condition:
\begin{equation}
\epsilon \geq 0~, ~~~~~~ - \epsilon \leq p_z \leq \epsilon~, ~~~~~~ - \epsilon \leq p_\theta \leq \epsilon~,
\end{equation}
and the strong energy condition:
\begin{equation}
\epsilon + p_z + p_\theta \geq 0~.
\end{equation}

In summary the new exact solution to the cylindrically symmetric Einstein-Maxwell equations was obtained. The solution is singular on the axis of symmetry and at the infinity, where sources should be placed. Since our solution (\ref{Solution}) contains electromagnetic radiation and non-diagonal metric components, and since core region exhibits gravitational repulsion, the source at the origin can be interpreted as a charged domain wall shell.

\section*{Acknowledgments}

I acknowledge the support of a 2008-2009 Fulbright Fellowship.


\end{document}